\documentclass[conference]{IEEEtran}
\IEEEoverridecommandlockouts

\usepackage{cite}
\usepackage{amsmath,amssymb,amsfonts}
\usepackage{graphicx}
\usepackage{textcomp}
\usepackage{xcolor}

\usepackage{url}
\usepackage{amsmath}
\usepackage{amssymb}
\usepackage{graphicx}
\usepackage{subcaption}
\usepackage{booktabs} 

\usepackage{algorithm}
\usepackage{algpseudocode}
\usepackage{amsmath}
\usepackage{cite}

\def\BibTeX{{\rm B\kern-.05em{\sc i\kern-.025em b}\kern-.08em
    T\kern-.1667em\lower.7ex\hbox{E}\kern-.125emX}}
\begin{document}

\title{Llasa+: Free Lunch for Accelerated and Streaming Llama-Based Speech Synthesis\\

}

\author{

\IEEEauthorblockN{Wenjie Tian}
\IEEEauthorblockA{\textit{Northwestern Polytechnical University} \\
Xi'an, China \\
twj@mail.nwpu.edu.cn
}
\vspace{10pt}  

\hfill%
\IEEEauthorblockN{Zhen Ye}
\IEEEauthorblockA{\textit{Hong Kong University of  } \\
\textit{Science and Technology}\\
Hong Kong, China \\
zhenye213@gmail.com
}

\and
\IEEEauthorblockN{Xinfa Zhu}
\IEEEauthorblockA{\textit{Northwestern Polytechnical University} \\
Xi'an, China \\
xfzhu@mail.nwpu.edu.cn
}
\hfill
\vspace{10pt}  

\IEEEauthorblockN{Wei Xue}
\IEEEauthorblockA{\textit{Hong Kong University of  } \\
\textit{Science and Technology}\\
Hong Kong, China \\
weixue@ust.hk
}

\and
\IEEEauthorblockN{Hanke Xie}
\IEEEauthorblockA{\textit{Northwestern Polytechnical University} \\
Xi'an, China \\
1490824861@mail.nwpu.edu.cn
}
\vspace{10pt}  

\hfill%
\IEEEauthorblockN{Lei Xie$^{*}$}
\IEEEauthorblockA{\textit{Northwestern Polytechnical University} \\
Xi'an, China \\
lxie@nwpu.edu.cn
}
\thanks{*Corresponding author.}
}

\maketitle

\begin{abstract}

Recent progress in text-to-speech (TTS) has achieved impressive naturalness and flexibility, especially with the development of large language model (LLM)-based approaches. However, existing autoregressive (AR) structures and large-scale models, such as Llasa, still face significant challenges in inference latency and streaming synthesis. 
To deal with the limitations, we introduce Llasa+, an accelerated and streaming TTS model built on Llasa. 
Specifically, to accelerate the generation process, we introduce two plug-and-play Multi-Token Prediction (MTP) modules following the frozen backbone. These modules allow the model to predict multiple tokens in one AR step.
Additionally, to mitigate potential error propagation caused by inaccurate MTP, we design a novel verification algorithm that leverages the frozen backbone to validate the generated tokens, thus allowing Llasa+ to achieve speedup without sacrificing generation quality.
Furthermore, we design a causal decoder that enables streaming speech reconstruction from tokens.
Extensive experiments show that Llasa+ achieves a 1.48× speedup without sacrificing generation quality, despite being trained only on LibriTTS. 
Moreover, the MTP-and-verification framework can be applied to accelerate any LLM-based model.
All codes and models are publicly available at~\url{https://github.com/ASLP-lab/LLaSA_Plus}.

\end{abstract}


\section{Introduction}

\begin{figure*}[]

  \centering
  \includegraphics[width=0.98\linewidth]{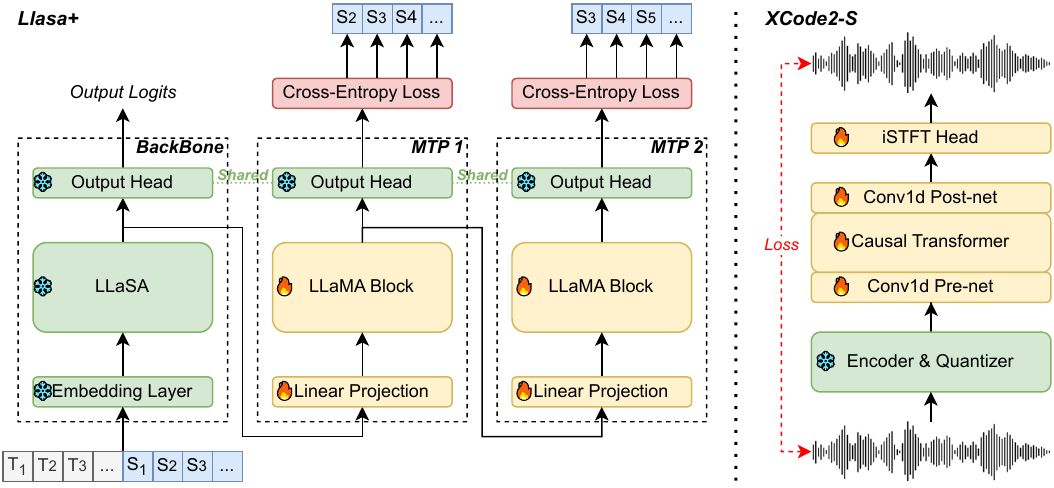}
\caption{Left: The architecture of Llasa+. Llasa+ consists of a frozen Llasa model and two MTP modules. MTP1 and MTP2 share the frozen LM head from Llasa, reducing modeling complexity during training. The backbone and MTP modules are organized in a cascaded structure, where the last hidden states from the previous module serve as input to the next. This carefully designed architecture enables the MTP modules to be plug-and-play and applicable to any LM-based model. During training, only the cross-entropy loss of the MTP modules is computed.
Right: The architecture of XCodec2-Streaming. The transformer in XCodec2 is modified into a causal transformer to support streaming generation. During fine-tuning, the decoder part is unfrozen.
}

\label{pipeline}
\end{figure*}

In recent years, with well-designed modules, larger datasets, and increased model size, text-to-speech (TTS) has made great progress in naturalness and quality. Represented by the language models (LM) and diffusion models~\cite{f5tts, CosyVoice2, seedtts, llasa, megatts2, wang2025spark, guo2024fireredtts, vocalnet}, TTS models are capable of synthesizing speech for any speaker by imitating the timbre, prosody and style of a reference speech. Among these models, Llasa~\cite{llasa}, an open-source and simplified TTS system, employs a single-layer vector quantizer (VQ) codec and a single Transformer architecture to fully align with standard LLMs such as Llama~\cite{llama}. Furthermore, through scaling train-time and inference-time compute, Llasa surpasses advanced TTS systems such as SeedTTS~\cite{seedtts} and CosyVoice~\cite{CosyVoice}, making it a promising approach for TTS tasks.



As large language model (LLM)-based TTS models become more powerful and flexible, they are increasingly integrated into real-time interactive applications.
Recent advances in dialogue systems~\cite{moshi, baichuan,freeomni, glm4, miniomni2, osum, kimiaudio, vocalnet, qwen25omni}, have demonstrated impressive capabilities in real-time human-computer interaction. In such interactive systems, streaming speech synthesis is a critical component. However, their autoregressive (AR) structure inherently limits inference speed, especially as the model size increases.
Therefore, many recent approaches are designed to further accelerate inference speed.
For example, CosyVoice2~\cite{CosyVoice2} introduces a chunk-aware causal flow matching model and a pre-trained vocoder to generate waveforms from speech tokens. Although the complicated design improves speed to some extent, it does not address the fundamental bottleneck: the slow autoregressive token prediction process.
Alternative approaches have been proposed to mitigate this issue. VALL-E 2~\cite{valle2} proposes Grouped Code Modeling, where multiple Transformer heads are used to predict groups of tokens at once.
However, this method requires retraining the backbone on a large-scale dataset to preserve its original performance. 
Another approach, proposed by Wang et al.~\cite{vibit}, leverages a Viterbi-like algorithm to speed up inference. However, its performance relies on a transition matrix learned from the training data, making it sensitive to data distribution shifts. 

In this paper, we propose Llasa+, an accelerated and streaming text-to-speech (TTS) model designed to improve AR inference efficiency while maintaining generation quality. Llasa+ consists of a frozen backbone model and two trainable Multi-Token Prediction (MTP) modules. Given Llasa's~\cite{llasa} strong performance in TTS, Llasa+ adopts it as the backbone to map input text tokens to speech logits. 
First, to enhance inference speed, inspired by DeepSeek-V3~\cite{deepseek}, Llasa+ incorporates two MTP modules that operate in sequence with the backbone. These MTP modules allow Llasa+ to predict multiple tokens in one AR step.
In addition, to mitigate potential error propagation caused by inaccurate MTP, a verification algorithm is employed to validate the predictions of the MTP modules. 
By accepting multiple verified tokens per inference step, Llasa+ successfully reduces the overall number of autoregressive iterations, resulting in a speedup of the generation process without compromising speech quality.
Finally, Xcodec2-Streaming (Xcodec2-S), the streaming speech token decoder, reconstructs high-fidelity speech waveforms streamingly from speech tokens. 
Adapted from the X-Codec2 architecture, Xcodec2-S adopts the causal decoder to focus solely on historical context, thus enabling streaming waveform reconstruction. 

Although Llasa+ is trained only on LibriTTS~\cite{libritts}, extensive experiments demonstrate its superior capability in streaming speech synthesis. Within the proposed MTP-and-verification framework, Llasa+ achieves 1.48× faster inference without performance degradation compared to Llasa.
To optimize the acceleration performance, we conduct comprehensive experiments exploring various architectural designs of the MTP module and hyperparameter configurations of the verification algorithm, ultimately identifying optimal settings.
Interestingly, in some configurations, the integration of the verification algorithm and MTP module not only accelerates generation but also leads to a modest improvement in speech quality beyond expectations.
Furthermore, with a lightweight yet effective modification, Xcodec2-S provides a practical and efficient solution for real-time TTS applications.

We open-source all code and models to facilitate future research. The project of Llasa+ is available at~\url{https://github.com/ASLP-lab/LLaSA_Plus}.

The key contributions of our work are summarized as follows:
\begin{itemize}
    \item We propose Llasa+, an open-source and accelerated streaming TTS model with two carefully designed MTP modules, achieving a 1.48× speed-up.
    \item We propose a novel plug-and-play MTP-and-verification framework that enables faster autoregressive inference without sacrificing generation quality and can be applicable to any LLM-based model.
    \item We introduce Xcodec2-S, a causal version of Xcodec2 providing an effective and efficient solution for streaming speech synthesis.

\end{itemize}

\section{Methodology}

Llasa+ is designed to support fast and streaming speech synthesis while retaining the generation quality of the original Llasa model. 
As shown in Fig.~\ref{pipeline}, building on the backbone, Llasa+ introduces two additional multi-token prediction (MTP) modules to perform multi-token prediction. Meanwhile, a novel verification mechanism is proposed to validate the generated speech tokens, preventing degradation of speech quality. Finally, Llasa+ incorporates a causal speech token decoder to support streaming speech reconstruction from predicted tokens.

\subsection{Multiple Token Prediction}
Inspired by DeepSeek-V3~\cite{deepseek}, each MTP block is composed of a linear projector followed by a LLaMA block.
Specifically, as illustrated in Fig.~\ref{pipeline}, since the last hidden states of the language model contain rich contextual information, we extract the last hidden states $ h_{0: t } ^0 $ produced by the backbone and sequentially process them through two MTP modules. This process can be formulated as follows:
{
\begin{equation}
\mathbf{h}_{0:t}^{k} = \text{MTP}_k(\mathbf{h}_{0:t}^{k-1})
\label{formu1}
\end{equation}
}
\noindent
where $ k $ represents the hidden state output of the $ k $-th MTP module, with $ k \in \{1, 2, \ldots, N - 1\} $. And $ 0:t $ denotes the input sequence of tokens from time step 0 to $ t $.

According to Equation~\eqref{formu1}, when $ k = 1 $, $ \mathbf{h}_{0:t}^0 $ is used as the input to the first MTP module~(MTP-1), from which we obtain the last hidden states of $ \mathbf{h}_{0:t}^1 $. Similarly, $ \mathbf{h}_{0:t}^1 $ is used as the input to the MTP-2, which adopts the same architecture as MTP-1, and the corresponding output is $ \mathbf{h}_{0:t}^2 $.
The results of these two hidden states, $ \mathbf{h}_{0:t}^1 $ and $ \mathbf{h}_{0:t}^2 $, are then fed into the LM head to produce token predictions:
   $ S_{1: t + 1} $ and $ S_{2:t + 2} $.
It is worth noting that, to maintain the performance while reducing training difficulty, we freeze the backbone and ensure that all MTP modules share the frozen LM head. Finally, the results $S$ produced by each MTP module are used to compute a cross-entropy (CE) loss with the ground-truth (GT) speech tokens $G$. Therefore, the total loss can be formulated as:
\begin{equation}
\mathcal{L}_{MTP} = \sum_{k=1}^{N-1} \, \mathcal{L}_{\text{CE}}(S_{0:T-k-1}, G_{k+1:})
\end{equation}
\noindent
where $ T $ denotes the total sequence length.
The target $G_{k+1:}$ is offset by $k+1$ steps to match the MTP module's goal of predicting the $(k+1)$-th future token.

\subsection{Verification Algorithm}

\newcommand{\COMMENTLLAMA}[1]{{\color{blue} $\triangleright$ {#1}}}
\begin{algorithm}[tb]
\caption{MTP-and-verification framework}
\label{alg:mtp_inference}
\footnotesize
\begin{algorithmic}[1]
\Require Initial input tokens $ x $, max generation length $ T $,  sampling hyperparameters $ h_{sample} $,
 verification hyperparameters $ topk\_v $ 

\Ensure Generated sequence $ S $.

\State \COMMENTLLAMA{Initialize the input sequence}
\State $ S \gets x $  
\State \COMMENTLLAMA{Tokens to be validated}
\State $ \textit{mtp1\_token}, \textit{mtp2\_token} \gets \texttt{None}, \texttt{None} $  

\For{$ \textit{step} = 1 $ to $ T $}
    \State \COMMENTLLAMA{Get logits from backbone model}
    \State $ \textit{logits} \gets \texttt{Llasa}(S) $  

    \State \COMMENTLLAMA{Verify \textit{mtp1\_token}}
    \If {$ \textit{mtp1\_token} \neq \texttt{None} $}   
    
        \If {$ \textit{mtp1\_token} \notin \texttt{Sample}(\textit{logits}[-3],  topk\_v ) $}
            \State $ \textit{new\_token} \gets \texttt{Sample}(\textit{logits}[-3],  h_{sample} ) $
            \State $ S \gets \texttt{ReplaceLastTwoTokens}(S[:-2], \textit{new\_token}) $
            \State $ \textit{mtp1\_token}, \textit{mtp2\_token} \gets \texttt{None}, \texttt{None} $
            \State \textbf{continue}
        \EndIf
        
        \State $ \textit{mtp1\_token} \gets \texttt{None} $

    \EndIf

    \State \COMMENTLLAMA{Verify \textit{mtp2\_token}}
    \If {$ \textit{mtp2\_token} \neq \texttt{None} $}   
    
        \If {$ \textit{mtp2\_token} \notin \texttt{Top-}k(\textit{logits}[-3],  topk\_v ) $}
            \State $ \textit{new\_token} \gets \texttt{Sample}(\textit{logits}[-2], h_{sample}) $
            \State $ S \gets \texttt{ReplaceLastToken}(S[:-1], \textit{new\_token}) $
            \State $ \textit{mtp2\_token} \gets \texttt{None} $
            \State \textbf{continue}

        \EndIf

        \State $ \textit{mtp2\_token} \gets \texttt{None} $

    \EndIf

    \State \COMMENTLLAMA{Sample token from backbone and MTP}

    \State $ \textit{token\_backbone} \gets \texttt{Sample}(\textit{logits}[-1], h_{sample}) $
    \State $ S \gets \texttt{Append}(S, \textit{token\_backbone}) $

    \State $ \textit{mtp1\_token} \gets \texttt{MTP1Predict}(\textit{hidden\_state}, h_{sample}) $
    \State $ S \gets \texttt{Append}(S, \textit{mtp1\_token}) $

    \State $ \textit{mtp2\_token} \gets \texttt{MTP2Predict}(\textit{hidden\_state\_from\_mtp1}, h_{sample}) $
    \State $ S \gets \texttt{Append}(S, \textit{mtp2\_token}) $
    
\EndFor

\State \Return $ S $

\renewcommand{\algorithmicrequire}{\textbf{Termination:} when $ \texttt{eos} \in S \land \textit{mtp1\_token} = \texttt{None} \land \textit{mtp2\_token} = \texttt{None} $ } 
\Require

\end{algorithmic}
\end{algorithm}

Due to the limited capacity of MTP modules, tokens predicted by MTP modules are not always accurate, leading to degraded synthesis performance.
To address prediction inaccuracy caused by MTP modules, we propose a novel verification algorithm that fully leverages the capabilities of the frozen backbone to validate the generated speech tokens, thereby ensuring the quality of synthetic speech. The detailed procedure of the algorithm is shown in Algorithm~\ref{alg:mtp_inference}.

Specifically, we first assume that the generation capability of the backbone is reliable. At time step $t$, given the previously generated tokens $ S_{0:t-1} $, the model simultaneously predicts three tokens: $ S_t $, $ S_{t+1}' $, and $ S_{t+2}' $. Here, $ S_t $ is generated by the backbone model, and is therefore considered a "trusted" token. However, $ S_{t+1}' $ and $ S_{t+2}' $ are produced respectively by MTP1 and MTP2 modules, making them "untrusted" tokens. Although the two "untrusted" tokens remain to be validated, all three tokens are added to the generated token list.

At time step $ t+1 $, the backbone model produces new logits: $ \text{logits}_{t+1} $, $ \text{logits}_{t+2}' $, and $ \text{logits}_{t+3}' $. Among these, since $ \text{logits}_{t+1} $ is generated by the "trusted" token, $ S_t $, 
so we use $ \text{logits}_{t+1} $ to verify the previous untrusted token $ S_{t+1}' $. 
Specifically, we check whether $ S_{t+1}' $ appears in the top-$ k $ candidates of $ \text{logits}_{t+1} $. 
If not, both $ S_{t+1}' $ and $ S_{t+2}' $ are deemed unreliable and are removed from the generated token list. Subsequently, a new trusted token $ S_{t+1} $ is sampled from $ \text{logits}_{t+1} $ and appended to the token list for the next autoregressive (AR) prediction.
If $ S_{t+1}' $ is contained within the top-$ k $ predictions of $ \text{logits}_{t+1} $, then $ S_{t+1}' $ is considered a trusted token, and we set $ S_{t+1} = S_{t+1}' $. In this case, the corresponding logits $ \text{logits}_{t+2}' $ generated from $ S_{t+1}' $ can be used to verify $ S_{t+2}' $ in the same manner. If $ S_{t+2}' $ is among the top-$ k $ predictions of $ \text{logits}_{t+2}' $, it is also deemed trustworthy, and we set $ S_{t+2} = S_{t+2}' $. Otherwise, $ S_{t+2}' $ is removed from the generated token list.

It is worth noting that any end-of-sequence (EOS) token from MTP modules must be verified before being accepted.

\begin{figure*}[h]
  \centering
  \includegraphics[width=0.9\linewidth]{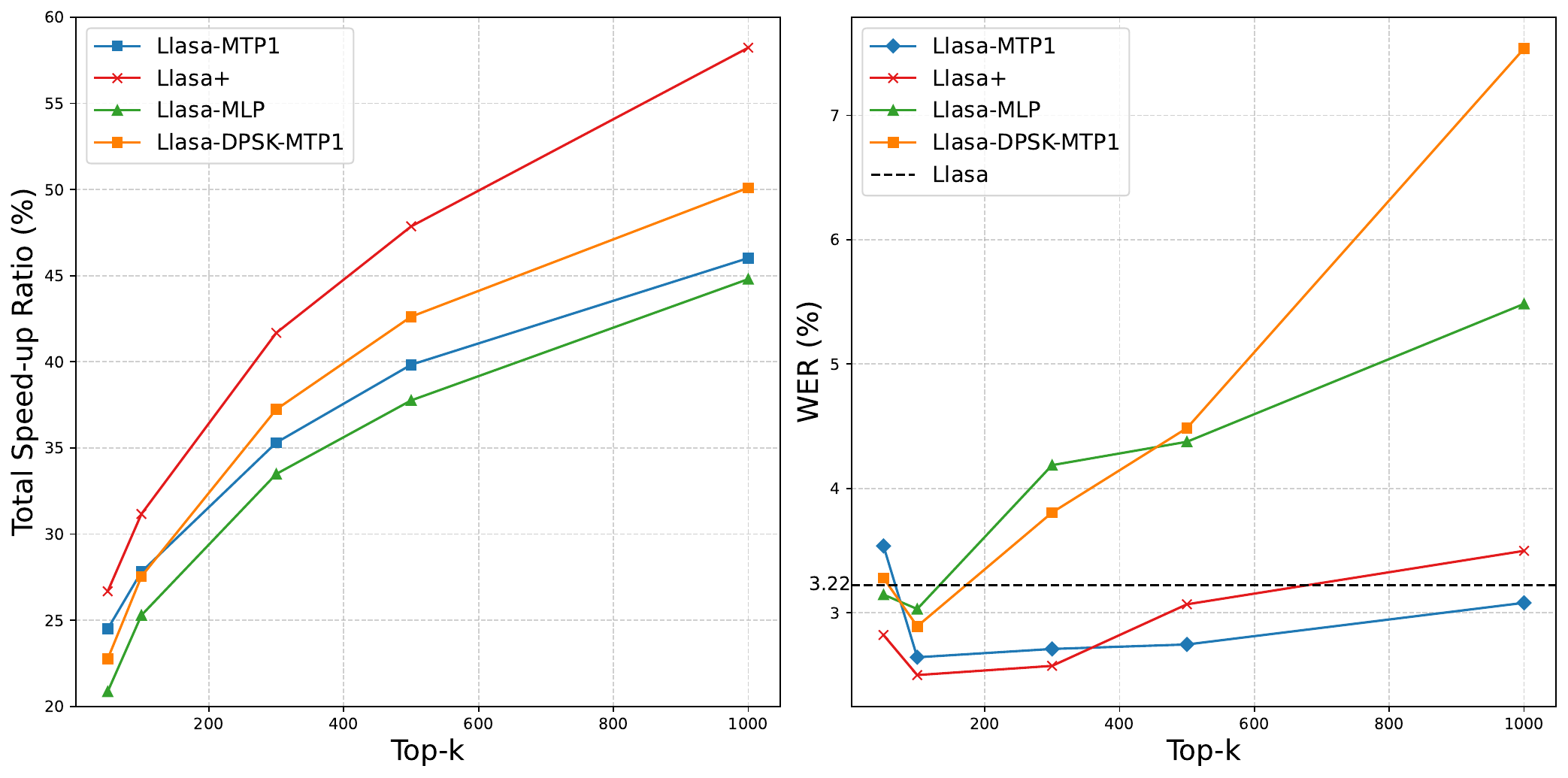}
\caption{The left part illustrates how the speed-up ratios of different variants change with the sampling parameter top-$k$.
The right part shows how the WER (Word Error Rate) of different variants varies as top-$k$ changes. The black dashed line represents the baseline's WER, directly adopted from Llasa.
The sampling parameter top-$k$ is varied over the set {50, 100, 300, 500, 1000}.
}
\label{fig:topk}
\vspace{-10pt}
\end{figure*}

\subsection{Streaming X-Codec2}

The Streaming X-Codec2 architecture consists of three main components: the encoder, the single vector quantizer (VQ) module, and the decoder.
Given a raw speech waveform $ Y $, the encoder maps it into a continuous latent representation, denoted as $ \mathbf{h} = \text{Encoder}(Y) $. This representation then serves as the input to the VQ module, which discretizes the speech representation into a sequence of codes.

To support streaming waveform reconstruction, the decoder employs a causal Transformer-based architecture that autoregressively predicts the short-time Fourier transform (STFT) magnitude and phase coefficients from speech tokens. These spectral predictions are subsequently converted into time-domain waveforms via an inverse STFT (iSTFT) head.
During training, both the encoder and the VQ module are initialized with pretrained weights from X-Codec2~\cite{llasa} and kept frozen and used solely for speech token extraction.
To enhance the quality of streaming decoding, the conv1d layers in the decoder are trainable as an adapter. 
Notably, while most of the streaming decoder operates in a strictly causal manner, 
Some conv1d layers utilize local lookahead mechanisms to slightly anticipate future context within a fixed window. This trade-off enhances synthesis quality without compromising streaming capabilities. 

\section{Experimental Setup}

\subsection{Datasets}

The training corpus for Llasa+, including both MTP modules and XCodec2-S, consists of the LibriTTS~\cite{libritts} training set: train-clean-100, train-clean-360, and train-other-500.

For evaluation, we adopt Seed-TTS-eval-en as the test set to assess the synthesis performance of Llasa+. Additionally, the LibriSpeech~\cite{librispeech} test-clean set is employed to evaluate the causal codec's modeling capability and speech reconstruction quality.

\subsection{Implementation Details}

Llasa+ is built upon Llasa-1B~\cite{llasa}, a model based on the LLaMA model~\cite{llama}. Llasa+ adopts 16 LLaMA decoder layers with a hidden size of 2048, 32 attention heads, and an 8192-dimensional feed-forward network (FFN). Each MTP block consists of a single LLaMA decoder layer with the same configuration.

The Streaming X-Codec2 adapter is implemented as a linear layer that maps from a 2048-dimensional input to a 1024-dimensional output.

The MTP blocks are trained on 4 $\times$ NVIDIA A800 GPUs with a total batch size of 256 for 20 epochs. The maximum learning rate is set to $1 \times 10^{-4}$, with 4000 warmup steps. A cosine learning rate scheduler with warmup is used, along with the AdamW optimizer and betas $(0.9, 0.999)$.
XCodec2-S is trained on 8 $\times$ NVIDIA 4090 GPUs with a total batch size of 96 for 280k steps. The maximum learning rate is set to $1.0 \times 10^{-4}$, with 3000 warmup steps. 
The same cosine learning rate scheduler with warmup is applied, together with the AdamW optimizer and betas $(0.8, 0.9)$. All other hyperparameters are kept consistent with those of X-Codec2.

\subsection{Comparison Models}
To comprehensively evaluate the effectiveness of our proposed method, we design a series of variants based on the Llasa framework for ablation studies.
Each system is listed as follows.

\begin{itemize}
    \item \textbf{Llasa-MTP1}: We further investigate structural modifications within Llasa. Llasa-MTP1 uses a single MTP block instead of multiple blocks as in Llasa+.
    \item \textbf{Llasa-DPSK-MTP2}: Following the approach of DeepSeek-V3~\cite{deepseek},  Llasa-DPSK-MTP2 employs the architecture where the input to MTP consists of not only the last hidden states of $ h_{0:t} $, but also the speech tokens of $ G_{1:t+1} $.
    \item \textbf{Llasa-DPSK-MTP1}: It uses a single MTP block compared to Llasa-DPSK-MTP2.
    \item \textbf{Llasa-MLP}: Compared to Llasa-MTP1, Llasa-MLP replaces the attention layers in the MTP module with MLP to assess the necessity of attention mechanisms. 
    \item \textbf{Llasa-Valle}: Following Valle2~\cite{valle2}, we also explore a Transformer-based LM head. For a direct comparison with MTP2, Llasa-Valle2 incorporates two additional independent decoder layers, resulting in the parallel prediction of three tokens at once.

\end{itemize}

\subsection{Evaluation Metrics}

The evaluation metrics used in this work are consistent with those of Llasa.

For the MTP module, we evaluate using the seed-tts-eval~\footnote{\url{https://github.com/BytedanceSpeech/seed-tts-eval}} toolkit. Speaker similarity (SIM) and Word Error Rate (WER) are adopted for objective evaluation.

For the codec component, we use the following metrics to assess both speech quality and speaker similarity.
a HuBERT-based ASR system is employed to compute the Word Error Rate (WER)~\footnote{\url{https://huggingface.co/facebook/hubert-large-ls960-ft}},
the Short-Time Objective Intelligibility (STOI) score~\footnote{\url{https://github.com/mpariente/pystoi}}, the Perceptual Evaluation of Speech Quality (PESQ) score~\footnote{\url{https://github.com/ludlows/python-pesq}},
and UTMOS~\footnote{\url{https://github.com/tarepan/SpeechMOS}} are used to assess speech quality. And a WavLM-based speaker verification model is utilized to measure Speaker Similarity (SPK-SIM)~\footnote{\url{https://github.com/microsoft/UniSpeech/tree/main/downstreams/speaker_verification}}.

\section{Experimental Results}

\subsection{Sampling-Parameter Topk-$k$ }

As shown in the left part of Fig.~\ref{fig:topk}, the model speedup ratio increases with top-$k$. Clearly, when top-$k$ increases, the acceptance rate of tokens predicted by MTP also increases, leading to a natural rise in speedup. 
Notably, when top-$k$ is set to 500, Llasa+ achieves the highest speedup without sacrificing model performance. Although Llasa-MTP1 can achieve a comparable speedup when top-$k$ is increased to 1000, its generation quality is inferior to that of Llasa+.

As shown in the right part of Fig.~\ref{fig:topk}, we observe that as the top-$k$ sampling size increases, the WER of variants generally decreases first and then increases. All of the variants achieve the best performance when top-$k$ is 100, significantly outperforming the baseline model in WER.
This improvement may be attributed to the high codec bitrate in TTS, and the MTP module enables the model to better leverage longer-term historical information, rather than focusing solely on the immediate past step.

\subsection{MTP Variants}
\label{MTP Variants}

The experimental results of MTP variants are shown in Table~\ref{table_mtp}. Equipped with the MTP-and-verification framework, the best models among Llasa-MTP1 and Llasa+ are even able to achieve considerable acceleration while simultaneously improving model performance. 
Specifically, the optimal Llasa-MTP1 configuration reduces the Word Error Rate (WER) from 3.220 to 2.642 and increases the similarity score (SIM) from 0.572 to 0.583, while achieving a $1.28\times$ acceleration in token prediction. Similarly, the best Llasa+ model achieves a WER reduction from 3.220 to 2.499 and a SIM improvement from 0.572 to 0.575, along with a $1.42\times$ speedup. If we relax the performance requirements and aim for parity with the backbone, Llasa+ can deliver more than $1.5\times$ speedup, as shown in Fig.~\ref{fig:topk}.
If top-$k$ is 500, Llasa+ achieves a $1.48\times$ speedup in token prediction while maintaining competitive performance.

Furthermore, under the same number of MTP modules and sampling parameters, both Llasa-DPSK-MTP1 and Llasa-DPSK-MTP2 exhibit inferior performance compared to their respective counterparts, Llasa-MTP1 and Llasa+. Incorporating ground-truth tokens as input to the MTP modules results in performance degradation. These results may be attributed to the discrepancy between training and inference conditions: during training, the input consists of ground-truth speech tokens, whereas during inference, the tokens are predicted and thus deviate from the ground truth. This mismatch likely exacerbates cascading errors and leads to a performance gap.

\begin{table}[]
\centering
\setlength{\tabcolsep}{3pt} 
\caption{Objective evaluation of MTP variants. 
Unless explicitly specified, the top-$k$ value for each variant defaults to 500. In the case of MTP2-based variants, the overall speed-up ratio is calculated as the sum of the first MTP module speed-up ratios and the second MTP module speed-up ratios. 
The best and second-best results are shown in \textbf{bold} and \underline{underlined}, respectively.}
\label{table_mtp}
\begin{tabular}{@{}l@{\hspace{4pt}}c|ccc@{}}
\toprule
Model             & Top-$k $   & WER(\%) ↓   & SIM ↑   & Speed-Up Ratio(\%) ↑ \\ \midrule
Llasa~\cite{llasa}   &   -  & 3.220  & 0.572 & 0                  \\
Llasa-Valle      &  500   & 23.565 & 0.436 & \textbf{100.00+100.00 (200.00)}  \\

Llasa-DPSK-MTP1     & 500    & 4.485 & 0.560 & 42.62              \\
Llasa-DPSK-MTP2   & 500    & 4.728 & 0.562 & \underline{32.74+16.52 (49.26)}              \\

Llasa-MLP     &  500   &   4.375 & 0.566 & 37.76              \\
Llasa-MTP1    &  100   & \underline{2.642} & \textbf{0.583} & 27.81    \\
Llasa-MTP1             & 500    & 2.745 & 0.571 & 39.83              \\
Llasa-MTP1    &  1000   & 3.080  & 0.562 & 46.02              \\

Llasa+         & 100    & \textbf{2.499} & \underline{0.575} & 28.92+12.76 (41.68)           \\ 
Llasa+              & 500    & 3.070  & 0.570 & 32.21+15.66 (47.87)  \\
\bottomrule
\end{tabular}
\vspace{-10pt}

\end{table}

It is worth noting that the acceleration ratio achieved by the second MTP module is significantly lower than that of the first MTP module. To further investigate the effect of increasing the number of MTP modules, we conduct an experiment with three MTP modules (MTP3). However, Llasa-MTP3 gains less than a 10\% acceleration, offering limited benefits while introducing additional costs, including bigger model size, increased training time, inference latency, and verification cost.
Therefore, using one or two MTP modules achieves a more favorable trade-off between performance and efficiency.

\begin{table*}[t]
\centering
\caption{
Experimental results on the ablation study of different XCodec2-S's components. The best and second-best results are shown in \textbf{bold} and \underline{underlined}, respectively.}
\label{table_xcodec}
\begin{tabular}{lcccccc}
\toprule
Model                                                                 & WER(\%) ↓  & STOI ↑  & PESQ\_WB ↑       & PESQ\_NB ↑ & SPK-SIM ↑ & UTMOS ↑          \\ 
\midrule
Ground Truth                                                          & 1.960 & 1.000  & 4.640           & 4.550     & 1.000    & 4.090           \\ 
\midrule
\begin{tabular}[c]{@{}l@{}}XCodec2~\cite{llasa}  \end{tabular}            & \textbf{2.470} & \textbf{0.919} & \textbf{2.433}          & \textbf{3.036}    & \textbf{0.821}   & \textbf{4.127}  \\     

XCodec2-S         & \underline{3.239}    & \underline{0.913}   & \underline{2.340} & \underline{2.932}    & \underline{0.795}   & \underline{4.029} \\

\quad w linear   &  3.446    & 0.911 & 2.321          & 2.931    & 0.793   & 4.023          \\
\quad w cov1d      &  3.971    & 0.912 & 2.315          & 2.921    & 0.793   & 4.028          \\ 
\bottomrule
\end{tabular}
\vspace{-10pt}
\end{table*}

In addition, we also conduct experiments on the MTP architecture to evaluate the impact of different decoder layer types. Specifically, based on the Llasa-MTP1, we compared decoder layers based on the attention mechanism with those utilizing multi-layer perceptrons (MLPs), while maintaining an equivalent number of parameters. As shown in Table~\ref{table_mtp}, replacing attention-based decoder layers with MLP-based ones leads to consistent performance degradation across all evaluation metrics. 
In particular, the WER increases significantly, rising from 2.745 to 4.375, while the similarity score drops by approximately 0.05. Moreover, the deterioration in speech token prediction performance also negatively impacts inference acceleration, with the processing speed decreasing from 39.83 to 37.76. 
These results highlight the critical role of attention mechanisms in enabling accurate sequential token prediction.

\subsection{XCodec2-S}

As shown in table~\ref{table_xcodec}, in the framework of Xcodec2-S, only the causal decoder is trainable, which preserves approximately 95\% of the original performance, demonstrating comparable results to the pretrained Xcodec2. 
We conducted two ablation studies to further investigate whether the streaming synthesis quality of Xcodec2-S can be improved by introducing additional trainable parameters or modifying the model architecture.

First, we add a trainable linear to Xcodec2-S, which lead to a degradation in performance, with the WER increasing from 3.239 to 3.446. Second, we modified the kernel size of the convolution layers in Xcodec2-S from 7 to 5. This adjustment resulted in a more significant drop in audio quality, with the WER rising to 3.971.
We hypothesize that these performance declines may stem from the relatively limited training data compared to the original pretrained model, which could adversely impact the codec’s generalization capacity.

\section{Ablation Study}

\begin{table}[t]
\centering
\caption{
Experimental results on the ablation study of verification algorithm. In the case of MTP2-based variants, the overall speed-up ratio is calculated as the sum of the first MTP module speed-up ratios and the second MTP module speed-up ratios. 
The best and second-best results are shown in \textbf{bold} and \underline{underlined}, respectively.
}
\label{table_ablation}
\begin{tabular}{lccc}
\toprule
Model                 & WER(\%) ↑   & SIM ↑   & Speed-Up Ratio(\%) ↑ \\ 
\midrule
Llasa+                  & \textbf{3.070}  & \textbf{0.570} & 32.21+15.66 (47.87)   \\   
\quad w/o Verification & 14.372 & 0.463  & \textbf{100.00+100.00 (200.00)}              \\ 
\quad w/o eos top-$k$    & \underline{3.422} & \textbf{0.570} & \underline{32.39+15.75 (48.14)}              \\ \midrule

Llasa-MTP1                  & \textbf{2.745} & \textbf{0.571} & 39.83             \\
\quad w/o Verification & 11.471 & 0.505 & \textbf{100.00}              \\
\quad w/o eos top-$k$    & \underline{3.319} & \underline{0.570} & \underline{40.33}              \\ 
\bottomrule

\end{tabular}
\end{table}

We conduct an ablation study to assess the effectiveness of the verification algorithm by evaluating its impact on speech generation. Results are shown in Table~\ref{table_ablation}.

\subsubsection{Verification Algorithm}

As presented in Table~\ref{table_ablation}, conducting MTP-$K$ alone without incorporating the corresponding verification algorithm results in substantial degradation in generation quality.

Specifically, for the configuration of Llasa+, the WER increases dramatically from 3.070 to 14.372, while the SIM score drops from 0.570 to 0.463. Such performance levels are unacceptable in practical text-to-speech (TTS) applications. The ablation results from both Llasa-MTP1 and Llasa+ clearly demonstrate the effectiveness and necessity of the verification algorithm. It enables acceleration without compromising the generation quality of the backbone, which is essential for real-world speech applications.

\subsubsection{Verification Hyperparameter}

During our experiments, we observe that omitting verification of the end-of-sentence (EOS) token resulted in worse performance. 
The model becomes more unstable and tends to terminate abnormally. To further investigate this, we perform an ablation study where the EOS check was not specially handled. Instead, its parameters are aligned with MTP sampling. As shown in Table~\ref{table_ablation}, both Llasa-MTP1 and Llasa+ drop in generation quality. Llasa+ w/o eos top-$k$ results in a significant increase in WER, rising from 3.070 to 3.422.
These results highlight the critical role of dedicated verification for the EOS token in maintaining model stability and overall performance.



\section{Conclusion}
\label{sec:con}

In this work, we present Llasa+, an accelerated and streaming model built upon the frozen text-to-speech (TTS) model Llasa. Equipped with a plug-and-play MTP module and a novel verification algorithm, Llasa+ enables faster autoregressive inference while maintaining high-quality speech generation.
We conduct extensive experiments on various MTP module architectures and the hyperparameters of the verification algorithm. Ultimately, with two carefully designed MTP modules trained on the LibriTTS dataset, Llasa+ achieves a $1.48\times$ speedup without sacrificing generation quality. These findings highlight the potential of the MTP-and-verification framework as a general acceleration strategy for LM-based models and Xcodec2-S as a practical solution for streaming decoding.

\clearpage
\bibliographystyle{IEEEtran}
\bibliography{mybib}

\end{document}